\documentclass [a4paper,12pt]{article}
\usepackage {graphicx,amsmath}
\begin{document}
\begin{center}
 {\Large\bf Einstein-Bohr controversy and theory of hidden variables }     \\[2mm]
            Milo\v{s} V. Lokaj\'{\i}\v{c}ek    
\footnote{e-mail: lokaj@fzu.cz} \\
Institute of Physics of the AS CR, v.v.i., 18221 Prague 8, Czech Republic  \\

\end{center}
\vspace{3mm}

{\bf Abstract } \\
It has been shown by us recently that Einstein was right in his controversy with Bohr or that the so called hidden-variable theory should be preferred to the Copenhagen quantum mechanics. In the following paper the corresponding arguments will be shortly repeated. The main attention will be then devoted to explaining main differences between these two quantum alternatives, differing in the access to the problem of chance, causality and locality of microscopic objects. The actual meaning of the mentioned hidden variables will be discussed, too, the essence of which remained practically unclear during all past discussions. It will be shown that the theory of hidden variables (or Schr\"{o}dinger equation alone) is able to represent the properties of the whole known physical reality. 
    

\section{Introduction}
\label{sec1}
The fundamental theoretical physics in the twentieth century was essentially influenced by the controversy between Einstein and Bohr and by its solution that was accepted by the then physical community. The decisive preference was given to the Copenhagen quantum mechanics (CQM), while the so called hidden-variable theory (HVT) was refused. Later (after Bell proposed his inequalities) the discussion about these two alternatives was renewed; the decision was expected to be done on the basis of experimental data. The corresponding experiments (violating the given inequalities) were then interpreted as the full attestation of previous conviction.   
 In the following we shall give a short review of the given story with the aim to show that the conclusion based on the mentioned experiments was derived on the basis of the assumption that was mistakenly interpreted; it has corresponded to classical physics and not to the HVT. The fact that the idea of hidden parameters was rather unclear in the past contributed also significantly to the given conclusion.

Both the quantum theoretical alternatives were based on the validity of Schr\"{o}dinger equation. However, some important assumptions were added in the case of CQM by Bohr \cite{bohr1}. This difference between their assumption bases has not been, however, practically mentioned in the past. It has been spoken always only about one common mathematical model and two different interpretations. And the preference has been generally given to the CQM even if by some authors (see, e.g., \cite{home}) the HVT was preferred. The difference between the assumption bases of these quantum alternatives has been discussed practically only in \cite{conc2} (see also \cite{ffp9,hradec}), and corresponding consequences have been derived. 
It is then possible to say that also quite new answers have been obtained to the 
main questions considered in the round table discussion \cite{cetto} and having remained practically without necessary responses.

We shall begin with the short story of the whole problem (see Sec. \ref{sec2}). In Sec. \ref{sec3} the difference between two assumption bases will be introduced and main consequences will be presented. The classical fundamental of the assumption on the basis of which Bell's inequalities were derived will be demonstrated in Sec. \ref{sec4}. The essence of the so called hidden parameters will be explained in Sec. \ref{sec5}. In Sec. \ref{sec6} the possibility of deciding between both the quantum alternatives on experimental basis will be discussed. Internal discrepancies existing in the CQM will be then presented in Sec. \ref{sec7}. And some concluding remarks will be introduced in Sec. \ref{sec8}.
   
\section{Short story of given controversy}
\label{sec2}
Einstein was introducing some arguments against the CQM from the very beginning. However, his objections formulated around 1931 were not quite clear and fully reasoned. Only the objections published in 1935 \cite{einst} represented the start of the important discussion that lasted practically the whole century. Bohr \cite{bohr} refused categorically Einstein's arguments and insisted upon that the CQM described actual properties of matter world, even if it involved the non-local connection of microscopic objects at macroscopic distances. 

Einstein and his collaborators preferred in principle the so called theory of hidden variables the interpretation of which was not very clear at that time. They proposed a thought experiment (EPR experiment) that should have demonstrated the preference of their ideas. However, the common opinion was influenced at that time mainly by the statement of von Neumann \cite{vonne} published in 1932 that the new theory of Bohr did not admit any hidden variables. The physical community preferred then Bohr's standpoint. Nobody was taking at that time into account the paper of G. Hermann \cite{gher} that the argument of von Neumann should have been denoted as a circle proof. 

The situation changed partially in 1952 when Bohm \cite{bohm} showed that a hidden variable was contained practically already in the simplest Schr\"{o}dinger equation even if its import was not very clear. We will go back to this problem in Sec. \ref{sec5}, where we will try to explain what was concealed actually behind the idea of hidden variables. The idea of Bohm was then followed by Bell \cite{bell} who admitted fully Einstein's alternative and derived in 1964 some inequalities that should have allowed (as it was believed at that time) to decide the given controversy on the basis of experimental data. 

The thought experiment of Einstein was modified to be possible to perform it. The actual experiment consisted then in the measurement of coincidence transmission probabilities of two photons having the same polarization and going in opposite directions through two polarizers. The corresponding experimental results were available practically in 1982 \cite{asp}. It was shown that Bell's inequalities were violated, which was interpreted as the refusal of the HVT; only the CQM seemed to be admitted. 

At that time it was commonly believed that the inequalities of Bell corresponded to the HVT. However, in their derivation a rather strong assumption was involved. And it has been possible to show now that it has corresponded to the conditions of classical physics. The violation of the given inequalities has meant that the HVT as well as the CQM must be taken as admitted on the basis of EPR experiment; only the classical alternative has been refused. 

The EPR experiments could not give, therefore, any decision between the two quantum alternatives and some new ways had to be looked for, as it has been summarized, e.g., in Ref. \cite{conc2}. Important points will be explained in the following to a greater detail.

\section {Different assumption bases of two quantum alternatives}
\label{sec3}
It has been already mentioned that the HVT is practically equivalent to the Schr\"{o}dinger equation, which may be characterized by two main assumptions: 
\begin{itemize}
\item {first, it is the validity of time-dependent Schr\"{o}dinger equation \cite{schr} 
 \begin{equation}
  i\hbar\frac{\partial}{\partial t}\psi(x,t)=H\psi(x,t), \;\;\;\;
     H=-\frac{\hbar^2}{2m}\triangle + V(x)   \label{schr}
\end{equation}
where Hamiltonian $H$ represents the total energy of a given physical system and $x$ represents the coordinates of all matter objects;} 
\item{ physical quantities are expressed as the expected values of corresponding operators:
 \begin{equation}
           A(t) = \int\psi^*(x,t)\,A_{op}\,\psi(x,t)dx    
\end{equation}
where $A_{op}$ and functions $\psi(x,t)$ are operators and vectors in a suitable Hilbert space.}
\end{itemize}  

In the CQM two following important assumptions were then added:
\begin{itemize}
\item {the corresponding Hilbert space was required to be spanned on one set of Hamiltonian eigenfunctions $\psi_E(x)$:
\begin{equation}
                H\psi_E(x) = E\psi_E(x);
\end{equation} }
\item {the mathematical superposition principle was interpreted in physical sense, i.e., any superposition of two vectors (or any vector of the given Hilbert space) represented "pure"  physical state.} 
\end{itemize}

Even if the first two assumptions are shared by both the quantum alternatives the properties of them are fundamentally different. While in the CQM the time evolution of any physical system is time reversible, the evolution in the HVT is irreversible if the third assumption is refused and the Hilbert space is suitably chosen (extended). 

 Let us consider the simplest physical system that may exhibit a measurable time evolution: a two-particle system. The corresponding Hamiltonian may possess continuous or discrete spectrum. 
In the case of continuous spectrum the Hilbert space $\mathcal H$ will consist of two mutually orthogonal subspaces corresponding to incoming and outgoing states:  
\begin{equation}
      \mathcal{H}  \;\equiv\;  \{\Delta^-\oplus\Delta^+\}
\end{equation}
that are mutually related by evolution operator 
\begin{equation}
            U(t) \;=\; e^{-iHt} \;\;  (t \ge 0);
\end{equation}
the evolution going always in one direction from $\Delta^-$ to $\Delta^+$.
 It holds then also 
\begin{equation}
\mathcal{H} \;=\; \overline{\Sigma_t U(t)\Delta^-}
                             \;=\;\overline{\Sigma_t U(-t)\Delta^+}.
\end{equation}
Individual subspaces $\Delta^-$ and $\Delta^+$ are then spanned always on one set of Hamiltonian eigenfunctions in usual way. In the case of discrete spectrum the Hilbert space will consist of a chain of such subspace pairs (for details see \cite{kulo1,kulo2}).

As to the last assumption the consequences following from the function $\psi(x,t)$ has been strongly modified (or rather deformed). Even if the function $\psi(x,t)$ (being uniquely determined by Schr\"{o}dinger equation) involves the causality the evolution in the CQM must be interpreted as fully probabilistic at any time instant;  
only the probability distribution of individual quantities may be predicted, not their actual values.

On the other hand the evolution in the HVT is to be interpreted as causal, similarly as in the classical physics. In such a case it is, however, necessary to distinguish between the basic states (determined always by one Hamiltonian eigenfunction and representing "pure" states) and superposition states that must be interpreted now as "mixed" states (see also Sec. \ref{sec5}). The basic states evolve fully in agreement with the states of classical systems \cite{adv}. 

Both the quantum alternatives represent, therefore, two quite different theories. They cannot be interpreted in any case as two mere different interpretations of one common mathematical model. 

\section{Important assumption in Bell's inequalities}
\label{sec4}
It has been mentioned in Sec. \ref{sec2} that the violation of Bell's inequalities was interpreted in 1982 as the victory of the CQM. It was believed since 1964 even if any more detailed analysis was not performed that the assumption needed for their derivation corresponded to the HVT. However, it has been shown (see, e.g., \cite{conc2} or \cite{lk98}) that the given assumption has corresponded only to the classical physics and it has not been possible to attribute it to any quantum alternative.

For completeness we shall describe now shortly the essence of the corresponding mistake. As already mentioned the given experiment has consisted in the measurement of coincidence transmission probabilities of two photons having opposite spins and running in opposite directions through two polarizers:
            \[   <---|^{\beta}---o---|^{\alpha}--->  \]
where $\alpha$ and $\beta$ are deviations of individual polarizer axes from a common zero position.
And Bell \cite{bell} has derived the following inequalities
\begin{equation}
             B = a_1b_1+a_2b_1+a_1b_2-a_2b_2 \leq 2    \label{ineq}
\end{equation}
holding for any four transmission probabilities $a_j, b_j \;(j=1,2)$ corresponding to two different orientations of both the polarizers  (4 different combinations). The inequalities (\ref{ineq}) have been then attributed to the HVT.

However, the given assumption has demanded for individual probabilities the properties that have corresponded to conditions required by the classical physics. Any probability in the given experiment has been defined in principle by three different parameters: position of the photon source, photon spin direction and polarizer orientation. However, the applied assumption has excluded the influence of two last free parameters (comp. \cite{lok98}). 
 The given inequalities were derived, of course, in other ways, too; see, e.g., Ref. \cite{clau}. Similar assumptions have been, however, involved in all these approaches (see, e.g., \cite{lk98}), even if the contrary has been stated. 
 
The given situation may be represented more clearly when the individual probabilities $a_j$ and $b_k$ are substituted by operators representing individual measurement acts and applied in single subspaces (corresponding to individual polarizers) of the whole Hilbert space 
\begin{equation}
  {\mathcal H}\;=\; {\mathcal H}_a \otimes {\mathcal H}_b.   \label{tens}
\end{equation}
It holds for the expectation values of these operators (see \cite{hill})
   $$ 0\;\leq\; |\langle a_j\rangle|, \;|\langle b_k\rangle|\; \leq \;1\, .  $$
And it is evident that Bell's operator defined by Eq. (\ref{ineq}) may exhibit different limits according to basic assumptions concerning individual operators. 

 According to chosen commutative relations three different limits may be obtained (see \cite{conc2}):
    \[ |B| \;\; \leq \;\;\;  2, \;\; 2\sqrt{2}\;\;\mathrm{and}\;\;2\sqrt{3}.  \]
The first limit corresponds to the classical case, when all operators  $a_j$ and $b_k$ commute mutually, i.e., if
     $$ [a_j,b_k]= 0 \;,\;\;\;\;\;[a_1,a_2]\;=\;[b_1,b_2]\;= 0. $$  
The second limit corresponds to the HVT, when only the operators belonging to different subspaces commute (no interaction at distance), i.e., if
 $$ [a_j,b_k]= 0\;\;\;\;\mathrm {and}\;\;\;[a_1,a_2]\neq 0,\;[b_1,b_2]\neq 0\,.   $$
And finally, the third limit corresponds to the case when also the operators from different Hilbert subspaces do not commute (the interaction at distance), i.e., if
    $$  [a_j,b_k]\neq 0\;, \;\;\;[a_1,a_2]\neq 0, \;\;[b_1,b_2]\neq 0\,.    $$

Only the classical alternative has been, therefore, excluded by experimental EPR data. As to the HVT it does not contradict the results of EPR experiments (obtained, e.g., by Aspect et al. \cite{asp}). It is, of course, also the CQM that has not been excluded.

 And new arguments must be looked for to decide which of the quantum alternatives represents better the description of microscopic physical reality. However, before discussing this problem we shall try to explain what should be understood under the term of hidden variables. 

\section {Hidden variables and Schr\"{o}dinger equation }
\label{sec5}
It has been already mentioned that the HVT is practically identical with the Schr\"{o}dinger equation. And one should ask what is to be understood under the term of hidden variables. Especially, it is necessary to answer the question whether any hidden variables exist in the solutions of this equation.

The problem of hidden variables in the Schr\"{o}dinger equation was discussed since the beginning of the CQM, earlier than their existence was refused by von Neumann \cite{vonne}. The discussion was renewed partially after D. Bohm \cite{bohm} showed that such a parameter existed already in the simplest Schr\"{o}dinger equation. However, the actual existence of such parameters remained unclear.

To answer the question what is the actual essence of "hidden variables" one must ask what is the assumption basis on the grounds of which the Schr\"{o}dinger equation may be derived. And it has been shown by Hoyer \cite{hoyer} and Ioannidou \cite{ioan} that it has been possible to derive it if some energy distribution of states characterized by the solutions of Hamilton equations has been considered together with these equations. The basic solutions of Schr\"{o}dinger equation (characterized always by one Hamiltonian eigenfunction only) correspond then to individual solutions of the Hamilton equations and represent the so called "pure" states while the superpositions of these basic states must be interpreted always as "mixed" states \cite{adv}. 

Any superposition involves then a set of basic states corresponding to different energy values and evolving differently with time; the energy values of individual states being, however, saved for any basic state separately. It means that the basic states are not defined by the given energy values (many different states correspond to individual energy values). And as at any time instant the kinetic part of energy is defined by momentum values the given superposition state must correspond to a statistical distribution of the coordinates of individual objects or of the distances between these objects at a suitably chosen time instant (to define corresponding potential energy).   

The factual physical sense is fully hidden when one takes a general probabilistic distribution of basic states. However, the Schr\"{o}dinger equation allows to obtain directly results in concrete physical cases when, e.g., the results of collision experiments are studied. Here one starts practically always from one initial two-particle incoming state that may change into an outgoing state consisting of two or more particles. And the mentioned potential energy in any initial state may be characterized by the impact parameter value of two colliding particles, which is not known in individual cases, but only the statistical distribution of which may be at least approximately estimated. 
  
And this statistical distribution of impact parameter characterizes the given superposition. Consequently, in actual physical cases there are not in principle any "hidden" parameters; only some physical parameters determining the energy distribution in a superposition must be taken as statistically distributed.


\section {Two quantum theories and experimental data }
\label{sec6}
 The Copenhagen alternative has been often denoted as supported by different experimental data. However, in all such cases only the assumptions corresponding to the Schr\"{o}dinger equation (i.e., to the HVT) have been practically tested; without the last two assumptions (forming Copenhagen alternative) having been actually involved. And therefore, none of two quantum alternatives may be excluded on the basis of the experiments available in the past. 

At this place it is, of course, necessary to mention the statement of Belinfante contained in his book \cite{belin}. He argued that the CQM and the HVT had to give mutually different predictions in the case of EPR experiment, or in other words, that the prediction of HVT had to differ significantly from the Malus law (the approximately measured dependence of light transmission in the case of two polarizers, including the coincidence EPR experiment). This argument contributed surely significantly to giving the preference to the CQM in 1982 when practically the Malus law was obtained in the EPR experiment.

 The prediction of Belinfante was not, however, true as the given statement was based on mistaking interchange of transmission probabilities through a polarizer pair and one polarizer; a more detailed explanation having been given, e.g., in Refs. \cite{lk02,contr}. In fact, the approximate Malus law (as measured) has been fully consistent with the HVT.

Having discovered this discrepancy it was quite natural to ask: When in the EPR experiment the same predictions may be obtained for the so different physical concepts, would it not be possible to find an experiment where the predictions would be different? And after a preliminary theoretical analysis the measurement of light transmission through three polarizers has seemed to be a suitable way.

The corresponding experiments have been performed by us and the results have been obtained being fundamentally different from the predictions of the CQM; see Refs. \cite{krasa1,krasa2}. The comparative graphs with main results may be found in \cite{conc2}-\cite{hradec}. They have opened also a way for a more realistic interpretation of polarization phenomena, representing a complex polarization process as consisting practically of three effects: the influence of anterior and hinder boundaries and of proper polarizer medium. The given problem has been studied further by us; however, definite results have not yet been gained.

The given experimental data have preferred significantly the HVT to the CQM. However, the preference follows also from the fact that while any internal discrepancy cannot be found in the case of HVT several discrepancies exist for the CQM, which will be demonstrated in the following section.
 
\section{Internal discrepancies in CQM}
\label{sec7}
It is well known that the CQM accredits to the physical reality some properties that are denoted as logical paradoxes (wave-particle duality, non-local connection (entanglement), a.s.o.). However, these logical paradoxes follow from some contradictions contained in the corresponding mathematical model. And it is possible to introduce at least three contradictions from which mainly the first two may be denoted as decisive (see also \cite{conc2}):
\begin{itemize}
 \item{ The important discrepancy follows directly from the fourth assumption concerning the superposition principle. It concerns the existence of discrete states in Schr\"{o}dinger equation.  It is evident that if the last assumption has been added all mathematical superpositions should represent equivalent physical states (i.e., pure states) and no quantized (discrete) states can exist in principle in experimental reality. The given problem has been removed in the HVT when only eigenstates belonging to Hamiltonian eigenvalues may represent "pure" states and any superposition of theirs represents a statistical "mixture".}
 \item{ In the other discrepancy also the third assumption has played important role. Already in 1933 Pauli \cite{pauli} showed that it was necessary for the corresponding Hamiltonian to possess continuous energy spectrum from $-\infty$ to $+\infty$, which disagreed with the fact that the energy was defined as positive quantity, or at least limited always from below.   }
 \item{ In 1964 Susskind and Glogover \cite{suss} showed then that exponential phase operator   $\mathcal{E} \;=\;e^{-i\omega\Phi}$ (where $\Phi$ was the phase)  was not unitary, as it held
  $\;\mathcal{E}^\dag\mathcal{E}\:u_{1/2} \,\equiv\,  0$. 
It indicated that the given Hilbert space (defined according to the third assumption) was not complete to represent the evolution of a corresponding physical system quite regularly. } 
\end{itemize}
Many attempts have been done during the 20th century to remove the last two deficiencies. The reason of having been unsuccessful may be seen in the fact that practically in all these cases both the shortages were regarded and solved as one joint problem.  

The corresponding solution has been formulated only recently (see Refs. \cite{kulo1,kulo2}) when it has been shown that it is necessary to remove two mentioned shortages one after the other. The criticism of Pauli may be removed if the Hilbert space is extended in the accord to the physical system. E.g., as to the simple system of two free colliding particles it has had to be doubled against the third assumption as proposed by Lax and Phillips already in 1967 (see \cite{lax1,lax2}). It consists at least of two mutually orthogonal subspaces ($\mathcal{H} = \Delta^- \oplus \Delta^+$); each of them being spanned on one basis of Hamiltonian eigenfunctions (see Sec. \ref{sec3}). 
 
 As to the criticism of Susskind and Glogower the Hilbert space (extended already to solve Pauli's problem) should be further doubled and formed by combining two mutually orthogonal subspaces corresponding to systems with opposite angular momentums. 
These subspaces must be bound together by adding the action of exponential phase operator to link together the corresponding vacuum states $u_{1/2}$  as it was proposed already by Fain \cite{fajn}; see also \cite{kulo1,kulo2}. 
       
\section{Concluding remarks}
\label{sec8}
It follows from the preceding that Einstein was right in the controversy with Bohr and consequently one can derive that the Schr\"{o}dinger equation may be applied to the description of physical systems of any dimensions (microscopic as well as macroscopic) while the Copenhagen quantum mechanics must be refused on theoretical as well as experimental grounds (i.e., on the basis of internal discrepancies as well as of experimental data). The evolution of individual physical processes may be represented in the Hilbert space that is correspondingly chosen. It means, too, that there is not any gap between the descriptions of microscopic and macroscopic worlds. There is not more any reason, either, to refuse causality in the evolution of physical systems and the locality of microscopic objects.

In the HVT the so called "pure" states are represented always only by basic states (characterized always by one Hamiltonian eigenfunction) while the superposition states correspond directly to data obtained in concrete experiments (e.g., in collisions of two particles being characterized by the statistical distribution of impact parameter values).  

The total Hilbert space consists then of mutually orthogonal subspaces and its complete vector basis (formed always by Hamiltonian eigenfunctions only) represents all possible physical (classical or pure) states. And the physical properties of any superposition are given as the probabilistic sum of mutually orthogonal eigenstates (classical states), as the result depends always only on absolute values of complex coefficients in superposition amplitudes.

In the Copenhagen quantum mechanics two assumptions have been added that have deformed the original solutions of Schr\"{o}dinger equation in disagreement to reality. If the Hilbert space has been spanned on one set of Hamiltonian eigenfunctions only, the time evolution has been modified and the earlier causality (and time irreversibility) has changed; only probability distributions of physical quantities having been predicted. 

In the past it has been spoken about "hidden" variables in both the quantum alternatives. However, there are not in principle any hidden parameters in the Schr\"{o}dinger equation. These parameters may be identified with the physical parameters (i.e., coordinates or mutual distances) responsible for statistical distribution of basic states corresponding to individual energy values in given superpositions.
  
\end{document}